# Thermodynamic relations between Free Energy and Mobility

Andrew Boshi Li, Talid Sinno

## Abstract

Stochastic and dynamical processes lie at the heart of all physical, chemical, and biological systems. However, kinetic and thermodynamic properties which characterize these processes have largely been treated separately as they can be obtained independently for many systems at thermodynamic equilibrium. In this work we demonstrate the existence of a class of relations between kinetic and thermodynamic factors which holds even in the hydrodynamic limit, and which must be satisfied for all systems that obey detailed balance. We achieve this by proving that for systems with inhomogeneous equilibrium states governed by gradient flow dynamics such as the Cahn-Hilliard (CH) dynamics, the chemical potential and self-diffusivity must mutually constrain each other. We discuss common issues in the literature which result in inconsistent formulations, construct the consistency requirement mathematically, develop a class of self-diffusivities that guarantee consistency, and discuss the requirement originates from detailed balance and Boltzmann distribution.

The hydrodynamic limit of many physical processes can be understood in terms of systems of PDEs, which have the general structure of

$$\frac{\partial \phi_i}{\partial t} = L_i\big[\{M_1, M_2, \dots\}, F[\phi_1, \phi_2, \dots]\big], \qquad 1$$

where $L_i$ is a differential operator, $M_i = M_i[\{\phi_j\}]$ is the mobility of field $\phi_i$, $\{\phi_i\}$ is the set of scalar fields evolving in time, and $F[\{\phi_i\}]$ is a thermodynamic free energy functional. A specific instance of this, also known as the $H^{-1}$ gradient flow [1], is equivalent to the phenomenological continuum diffusion equation [2],

$$\frac{\partial c}{\partial t} = \boldsymbol{\nabla} \cdot D(c) \boldsymbol{\nabla} \mu, \qquad 2$$

where $D(c)$ is the self-diffusivity coefficient, $\mu[c] = \frac{\delta F}{\delta c}$ is the variational derivative of the free energy functional, $c$ is the concentration, and any temperature dependence is suppressed for notational simplicity.

Now consider that for any function(al)s $D(c)$ and $\mu[c]$, there always exists an effective chemical potential function(al), $\mu_D[c]$, which satisfies $\boldsymbol{\nabla} \mu_D \equiv D(c) \boldsymbol{\nabla} \mu$ such that

$$\frac{\partial c}{\partial t} = \nabla \cdot \nabla \mu_D = \nabla \cdot D(c)\nabla \mu. \qquad 3$$

Equation (3) highlights a constraint on $D(c)$ which arises from the fact that there are two ways to describe the equilibrium state of the system. The first equality states that the composite chemical potential, $\mu_D[c]$, completely determines the bulk phase behavior of the system. On the other hand, as we show later, equilibrium thermodynamics—namely the assumption of detailed balance and Boltzmann statistics—also requires that $\mu[c]$ alone fully specify the phase behavior. In other words, the assumptions behind equilibrium thermodynamics require that $\mu_D[c]$ and $\mu[c]$ predict the same equilibrium phases, which in turn induces a relationship between $\mu[c]$ and $D(c)$.

A mathematical statement of this constraint may be obtained by considering the common tangent construction of the binodal condition

$$\frac{G(c_a) - G(c_b)}{c_a - c_b} = \hat{\mu}(c_a) = \hat{\mu}(c_b), \qquad 4$$

where $c_a$ and $c_b$ are the binodal points. Here, $\hat{\mu}(c)$ represents the homogeneous chemical potential, i.e., $\mu[c] \approx \mu(c, (\nabla c)^2, \nabla^2 c, \ldots)$ and $\hat{\mu}(c) = \mu(c, 0, 0, \ldots)$, and $G$ is the homogeneous free energy, i.e., $\frac{\partial G}{\partial c} = \hat{\mu}$. Note that eq. (4) reflects the fact that only the homogeneous component of the chemical potential is responsible for setting the position of the binodal envelope. An equivalent condition for $\mu_D[c]$ is given by

$$\frac{G_D(c_b') - G_D(c_a')}{c_b' - c_a'} = \hat{\mu}_D(c_a') = \hat{\mu}_D(c_b'), \qquad 5$$

where $G_D$ is defined by $\frac{\partial G_D}{\partial c} = \hat{\mu}_D$. Consequently, eq. (3) and the assumption of equilibrium thermodynamics requires that

$$c_{a,b}' = c_{a,b}. \qquad 6$$

The constraints on $D(c)$ implied by eqs. (5) and (6) are then given by

$$\hat{\mu}_D(c_b) - \hat{\mu}_D(c_a) = \int_{c_a}^{c_b} D(c)\frac{\partial \hat{\mu}}{\partial c} dc = 0, \qquad 7$$

and

$$\frac{1}{c_b - c_a}\int_{c_a}^{c_b}\left(\int_0^{c'} D(c)\frac{\partial \hat{\mu}}{\partial c}dc\right)dc' = \int_0^{c_a} D(c)\frac{\partial \hat{\mu}}{\partial c}dc. \qquad 8$$

Analogous relations for non-conserved gradient dynamics (e.g., via the Allen-Cahn equation) are included in the SI.

We now illustrate how the constraints imposed by eqs. (7) and (8) are generally violated in practice by commonly assumed chemical potentials and diffusivities. Specifically, we consider a regular solution model,

$$\hat{\mu}(c) = \ln(c) - \ln(1 - c) + \alpha(1 - 2c), \qquad 9$$

and diffusivity of the form $D(c) = D_0 c(1 - c)$, which, when substituted along with eq. (9), into eq. (2) gives

$$\frac{\partial c}{\partial t} = \nabla \cdot D_0 c(1 - c)[\nabla(\ln(c) - \ln(1 - c) + \alpha(1 - 2c)) + \kappa\phi], \qquad 10$$

where $\phi$ represents all gradient contributions in $\mu(c, (\nabla c)^2, \nabla^2 c, ...)$ and $\kappa$ is a scaling parameter such that $k \to 0$ corresponds to the sharp interface limit [3]. Eq. (10) is widely employed as a workhorse model for describing diffusion in concentrated binary solutions, including spinodal decomposition [4], Li intercalation [5], and particle segregation [6]. Moreover, under certain conditions eq. (10) is the hydrodynamic limit of Kawasaki exchange dynamics on a lattice [6], and a series of papers on coarse-graining models for various exchange dynamics [7, 8] have shown that a broad class of microscopic diffusive processes are also well approximated in the continuum limit by eq. (10) and its variants. Next, following eq. (3), eq. (10) may be readily rearranged into the form

$$\frac{\partial c}{\partial t} = \nabla \cdot D_0(1 - 2\alpha c(1 - c))\nabla c = \nabla \cdot [D_0 \nabla(\hat{\mu}_D) + \kappa\phi], \qquad 11$$

where $\hat{\mu}_D = c - \alpha c^2 + 2\alpha c^3/3$. It is evident that while $\hat{\mu}$ and $\hat{\mu}_D$ predict the same spinodal points (i.e., $D(c)\frac{\partial \mu}{\partial c} = \frac{\partial \mu}{\partial c} = 0$), they predict different binodal points as shown in Fig. 1, violating the condition in eq. (6).

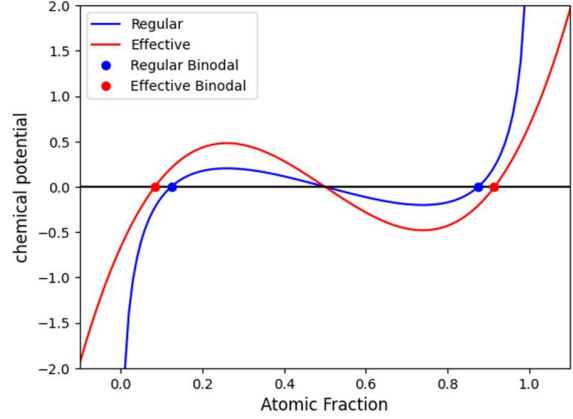

**Fig. 1**. Illustraion of how the binodal points predicted by $\hat{\mu}(c)$ and $\hat{\mu}_D$ differ. The blue curve corresponds to $\hat{\mu}(c)$, with the two blue dots corresponding to the predicted binodal points based on the chemical potential alone. The red curve corresponds to $\hat{\mu}_D$, with the two red dots corresponding to predictions of the binodal points based on the PDE. The latter is clearly displaced from the true binodals, which illustrates how eq. (10) can break down.

The question now remains—what, if any, are the allowable concentration-dependent diffusivities that guarantee binodal invariance while also accommodating direct experimental or simulation measurements [9]. Obviously, a constant diffusivity satisfies the binodal constraint but this is a poor description for many diffusive processes [10–12]. To show that in fact the binodal constraint is not as restrictive as it may initially seem, consider the class of homogeneous chemical potentials that is antisymmetric about $c = 0.5$ in the interval $c \in [0,1]$, i.e., $\hat{\mu}(0.5 + \delta c) = -\hat{\mu}(0.5 - \delta c)$. By the antisymmetric condition, the binodal compositions, $c_a$ and $c_b$, satisfy $\hat{\mu}(c_a) = \hat{\mu}(c_b) = 0$. For this class of chemical potentials, self-diffusivities of the form

$$D(\hat{\mu}) = \sum_n \alpha_n \hat{\mu}^{2n}, \qquad 12$$

where $n$ is a non-negative integer, and $\{\alpha_n\}$ are constant coefficients, guarantee binodal invariance. This is readily shown by noting that

$$D(\hat{\mu})\frac{\partial \hat{\mu}}{\partial c} = \frac{\partial \hat{\mu}}{\partial c}\sum_n \alpha_n \hat{\mu}^{2n} = \frac{\partial}{\partial c}\left(\sum_n \frac{\alpha_n}{2n+1}\hat{\mu}^{2n+1}\right) = \frac{\partial}{\partial c}\hat{\mu}_D, \qquad 13$$

which is still antisymmetric and satisfies $\hat{\mu}_D(c_a) = \hat{\mu}_D(c_b) = 0$. A practical example of the diffusivities represented by eq. (12) is

$$D_1(\hat{\mu}) = \frac{A}{\exp{-\mu} + \exp{\mu}}, \qquad 14$$

where $A$ is a fitting parameter. As shown in Fig. 2, for a regular solution model with $\alpha = 2.6$, setting $A = 0.085$, eq. (14) approaches $D = c(1-c)$ in the limits $c \to 1,0$ while enforcing binodal invariance and thus thermodynamic consistency.

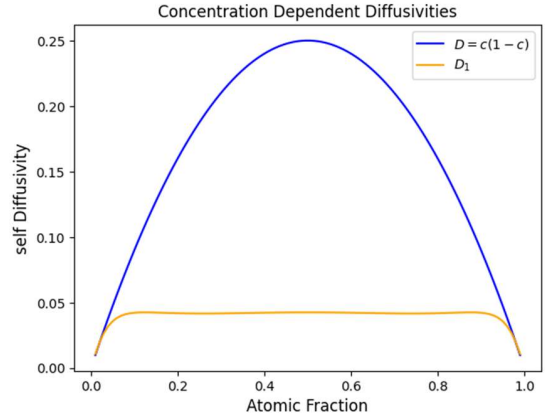

**Fig. 2**. An example of self-diffusivity in the class given by eq. (13) that satisfies binodal invariance for the regular solution model.

Finally, we investigate the microscopic origins of the binodal invariance constraint and its relationship to detailed balance and Boltzmann statistics. We begin with the master equation

$$\frac{\partial P[C]}{\partial t} = \sum (W[C',C]P[C'] - W[C,C']P[C]), \qquad 15$$

where $P[C]$ is the probability functional of state $C$, $W[C,C']$ is the transition rate from state $C$ to $C'$, $C$ is a compositional profile with $C = \{\ldots, c_i, \ldots, c_j, \ldots\}$, $C' = \{\ldots, c_i - \epsilon, \ldots, c_j + \epsilon, \ldots\}$, $\epsilon$ denotes the change in $c$ at site $i$. For any master equation satisfying eq. (15), the transition rate can be written as $W[C,C'] = \exp\{\beta\Omega[C,C']\}$, where $\Omega[C,C']$ is a functional of $C$ and $C'$ and is specified by an assumed kinetic model. With respect to the transitions from $C \to C'$ and $C' \to C$, there is a unique symmetric and antisymmetric decomposition in state space,

$$\Omega^{s/a}[C,C'] = \pm\Omega^{s/a}[C',C] = \frac{1}{2}(\Omega[C,C'] \pm \Omega[C',C]), \qquad 16$$

which, substituting into eq. (15), gives

$$\frac{\partial P[C]}{\partial t} = \sum \exp(\beta\Omega^s[C,C'])\,(\exp(\beta\Omega^a[C',C])\,P[C'] - \exp(\beta\Omega^a[C,C'])\,P[C]). \qquad 17$$

Next, we will relate eq. (15) to the diffusion equation following the approach of [13]. We generalize the arguments by showing how $\Omega^a$ relates to the chemical potential, $\Omega^s$ to a general mobility term and interpret the thermodynamic constraint on diffusivity in the context of stochastic processes.

We first discuss how $\Omega^a$ corresponds to a free energy difference of two states. Since various accounts of this relationship are used in literature with either no justification or simplified derivations, we provide an updated derivation and clarify all the underlying assumptions involved. We begin by invoking the detailed balance assumption, i.e.,

$$W[C,C']P_{eq}[C] = W[C',C]P_{eq}[C'], \qquad 18$$

where $P_{eq}[C]$ is the probability of the system being in state $C$ at equilibrium. We further assume that $P_{eq}[C]$ follows a Boltzmann distribution, i.e.,

$$P_{eq}[C] \propto \exp(-\beta F[C]), \qquad 19$$

where $F[C]$ is the free energy associated with state $C$. Substituting eq. (19) into eq. (18) gives

$$W[C,C']/W[C',C] = \exp(-\beta(F[C'] - F[C])), \qquad 20$$

or,

$$\Omega^a[C,C'] = \frac{1}{2}(F[C'] - F[C]). \qquad 21$$

The r.h.s. of eq. (21) can be expressed as the difference of free energy variational derivatives

$$\frac{1}{2}(F[C'] - F[C]) = \frac{1}{2}\int \frac{\delta F}{\delta c(\mathbf{r})}\delta c(\mathbf{r})dV + O((\delta c)^2) \approx \frac{1}{2}\epsilon\left(\frac{\delta F}{\delta c_j} - \frac{\delta F}{\delta c_i}\right), \qquad 22$$

or equivalently,

$$\Omega^a[C,C'] \approx \frac{1}{2}(\mu_j - \mu_i), \qquad 23$$

where the chemical potential $\mu_i \equiv \epsilon\frac{\delta F}{\delta c_i}$ is the change in total free energy due to addition of one atom at site $i$. In other words, assuming only detailed balance and Boltzmann distribution at equilibrium, we see that $\Omega^s[C, C']$ cannot alter the equilibrium distribution. This derivation also holds under a relaxed version of eq. (19) as long as $P_{eq}[C]$ only depends on $C$.

The final step is to establish a connection between the Master equation and the phenomenological diffusion model. Inserting eq. (23) into eq. (17) gives

$$\frac{\partial P[C]}{\partial t} = \sum_{i,j} \exp(\beta \Omega^s[C, C']) \left( \exp\left(\frac{\beta}{2}(\mu_j - \mu_i)\right) P[C'] - \exp\left(\frac{\beta}{2}(\mu_i - \mu_j)\right) P[C] \right). \qquad 24$$

This expression is related to a Langevin formulation of the form [XXXRisken]

$$\frac{\partial c_i}{\partial t} = \sum_j D_0 \exp(\beta \Omega^s[C, C'])(\mu_j - \mu_i) + \xi_i(t), \qquad 25$$

where $D_0$ is a constant prefactor, and $\xi_i(t)$ is a stochastic noise source [13]. Since the equilibrium state of the system is given by $\mu(\mathbf{r})$ alone, the choice of $\Omega^s[C, C']$ cannot affect the equilibrium profile. This is the microscopic origin of the constraint we have pointed out in the continuum diffusion equation as neglecting the noise term and assuming nearest neighbor exchanges only, in the limit of $\epsilon \to 0$, equivalence of eq. (1) and (24) gives

$$\log D(c) \propto \beta \Omega^s[C, C'] + \log(D_0), \qquad 28$$

and

$$\nabla \mu(c, \nabla^2 c) \propto \Omega^a. \qquad 29$$

Therefore, the thermodynamic constraint on diffusion originates from the symmetric and antisymmetric arguments of the corresponding transition rates of the master equation whenever detailed balance is satisfied, and the diffusion equation is the mean field approximation of the stochastic process.

Finally, having established the existence of thermo-kinetic relations in the case of gradient flows, we consider the more general class of constraints which arise from detailed balance. Assuming that the dynamics of a system is governed by free energy minimization and the bulk equilibrium phases are given by $\{\phi_i\}$. The rate at which the system evolves towards these minima, which is given by a set of mobilities $\{M_i\}$, cannot affect the equilibrium phases. This relation, which appears at first sight to be a statement about the independence of $\{M_i\}$ and $\{\phi_i\}$, can give rise to a whole range of relations between these two types of quantities. This can be understood graphically as shown in FIG. 3.

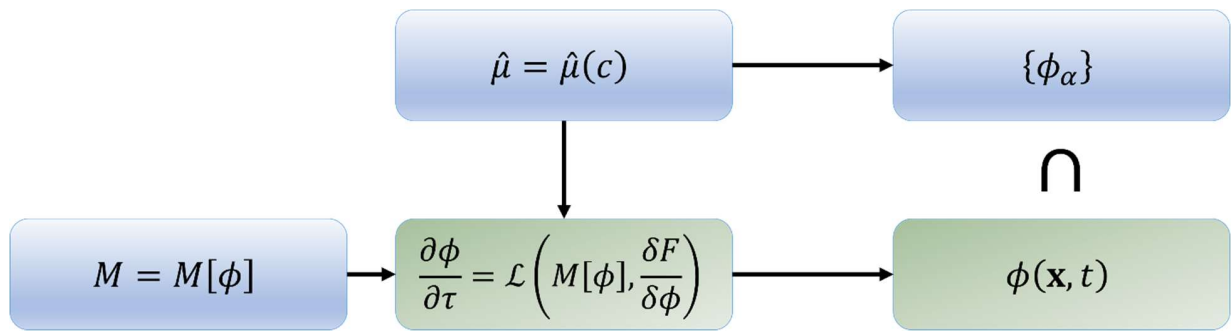

**Fig. 3**. A graphical depiction of how mobility and chemical potential is related. The relation holds for both conserved and non-conserved dynamics

While there has been great interest in establishing relations between diffusivity and free energy/entropy, such as empirical scaling laws proposed in the context of hard sphere and colloidal systems [14–17], there has been only limited success in their applicability. In this paper, we have demonstrated that it is necessary to develop a thermodynamically self-consistent formulation of self-diffusivity and argued that this can be understood in the dynamical context of the master equation and statistical mechanics. Furthermore, we provide a natural formulation of self-diffusivities which guarantee thermodynamic self-consistency by treating them as functions of chemical potential. Finally, by demonstrating that the relationship between mobility and chemical potential comes from the detailed balance and Boltzmann distribution assumptions, we point out that there exists a large class of relations that relate mobility and chemical potential (including but not limited to Cahn-Hilliard and Allen-Cahn dynamics)

# Bibliography

1. Li W, Bazant MZ, Zhu J (2023) Phase-Field DeepONet: Physics-informed deep operator neural network for fast simulations of pattern formation governed by gradient flows of free-energy functionals. *Computer methods in applied mechanics and engineering*, 416(116299):116299.

2. Cahn JW, Hilliard JE (1958) Free energy of a nonuniform system. I. interfacial free energy. *The Journal of chemical physics*, 28(2):258–267.

3. Elder KR, Grant M, Provatas N, Kosterlitz JM (2001) Sharp interface limits of phase-field models. *Physical review. E, Statistical, nonlinear, and soft matter physics*, 64(2 Pt 1):021604.

4. Zhao H, Storey BD, Braatz RD, Bazant MZ (2020) Learning the physics of pattern formation from images. *Physical review letters*, 124(6):060201.

5. Smith RB, Bazant MZ (2017) Multiphase porous electrode theory. *Journal of the Electrochemical Society*, 164(11):E3291–E3310.

6. Giacomin G, Lebowitz JL (1997) Phase Segregation in Dynamics in Particle Systems with Long Range Interactions. I. Macroscopic Limit. *J. Stat. Phys.*, 87

7. Katsoulakis MA, Vlachos DG (2003) Coarse-grained stochastic processes and kinetic Monte Carlo simulators for the diffusion of interacting particles. *The Journal of chemical physics*, 119(18):9412–9427.

8. Katsoulakis MA, Majda AJ, Vlachos DG (2003) Coarse-grained stochastic processes for microscopic lattice systems. *Proceedings of the National Academy of Sciences of the United States of America*, 100(3):782–787.

9. Chen C, Li WZ, Song YC, Weng LD, Zhang N (2012) Concentration dependence of water self-diffusion coefficients in dilute glycerol–water binary and glycerol–water–sodium chloride ternary solutions and the insights from hydrogen bonds. *Molecular physics*, 110(5):283–291.

10. Prokes SM, Wang KL (1990) Interdiffusion measurements in asymmetrically strained SiGe/Si superlattices. *Applied physics letters*, 56(26):2628–2630.

11. Xia G (maggie), Dong Y (2017) Si–Ge interdiffusion, dopant diffusion, and segregation in SiGe- and SiGe:C-based devices. *Micro- and Nanoelectronics*, :21–49.

12. Kaiser D, Ghosh S, Han S, Sinno T (2016) Multiscale Modeling of Stress-Mediated Compostional Patterning in SiGe Substrates. *ECS Tran*, 75(4):129–141.

13. Bronchart Q, Le Bouar Y, Finel A (2008) New coarse-grained derivation of a phase field model for precipitation. *Physical review letters*, 100(1):015702.

14. Sorkin B, Diamant H, Ariel G (2023) Universal relation between entropy and kinetics. *Physical review letters*, 131(14):147101.

15. Rosenfeld Y (1977) Relation between the transport coefficients and the internal entropy of simple systems. *Physical review A: General physics*, 15(6):2545–2549.

16. Rosenfeld Y (1999) A quasi-universal scaling law for atomic transport in simple fluids. *Journal of physics. Condensed matter: an Institute of Physics journal*, 11(28):5415–5427.

17. Dzugutov M (1996) A universal scaling law for atomic diffusion in condensed matter. *Nature*, 381(6578):137–139.